\documentclass[pdflatex,sn-mathphys]{sn-jnl}
\usepackage{graphicx}
\usepackage{amsmath}
\usepackage{amsfonts}
\usepackage{array}
\usepackage{siunitx}
\usepackage{bookmark}
\usepackage{natbib,hyperref}
\jyear{2022}
\begin{document}

\title{Magnetic cooling and vibration isolation of a sub-kHz mechanical resonator}

\author[1]{\fnm{Bernard} \sur{van Heck}}
\author[1]{\fnm{Tim M.} \sur{Fuchs}}
\author[1]{\fnm{Jaimy} \sur{Plugge}}

\author[2]{\fnm{Wim A.} \sur{Bosch}}

\author*[1]{\fnm{Tjerk H.} \sur{Oosterkamp}}\email{oosterkamp@physics.leidenuniv.nl}

\affil[1]{\orgdiv{Leiden Institute of Physics}, \orgname{Leiden University}, \orgaddress{P.O. Box 9504, 2300 RA Leiden, The Netherlands}}
\affil[2]{\orgname{HDL Hightech Development Leiden}, \orgaddress{Leiden, The Netherlands , hdl@freedom.nl}} 

\abstract{We report recent progress towards the realization of a sub-mK, low-vibration environment at the bottom stage of a dry dilution refrigerator for use in mechanical tests of quantum mechanics. Using adiabatic nuclear demagnetization, we have cooled a silicon cantilever force sensor to $T\approx 1$ mK. The temperature of the tip-holder of the cantilever chip was determined via a primary magnetic flux noise thermometer. The quality factor of the cantilever continues to increase with decreasing temperature, reaching $Q\approx 4\cdot 10^4$ at $2$ mK. To demonstrate that the vibration isolation is not compromised, we report the detection of the thermal motion of the cantilever down to $T \approx 20$ mK, only limited by the coupling to the SQUID readout circuit. We discuss feasible improvements that will allow us to probe unexplored regions of the parameter space of continuous spontaneous localization models.}

\maketitle

\section{Introduction}\label{sec1}

The preparation of a mechanical oscillator in a coherent quantum-mechanical state~\cite{poot2012} is one of the experimental frontiers~\cite{carlesso2022} that may test the limits, if any, of the applicability of quantum mechanics to the macroscopic world~\cite{leggett2002,arndt2014}.
Different approaches to this end, largely stemming from the field of optomechanics~\cite{aspelmeyer2014}, have been proposed and experimentally explored, with a wide range of physical systems of choice~\cite{poot2012}, which includes membranes~\cite{teufel2011}, nanobeams~\cite{chan2011}, cantilevers~\cite{marshall2003,kleckner2006,poggio2007},  levitated particles~\cite{delic2020,gonzalez2021}, dilatational resonators~\cite{oconnell2010}, and so on.
These systems span a wide range of frequencies, from kHz to GHz, and can be monitored not only via optical methods but also superconducting circuits~\cite{gely2021}.
Within this vast array of options, sub-kHz oscillators with a large mass, such as the soft magnet-on-tip silicon cantilevers used in magnetic resonance force microscopy (MRFM)~\cite{rugar2004}, are attractive to test and constrain~\cite{diosi2015,vinante2016} physical models of the wave function collapse~\cite{bassi2013}, particularly the continuous spontaneous localization (CSL) model~\cite{ghirardi1990} and the Di\'{o}si-Penrose model of gravitationally-induced collapse~\cite{diosi1987,penrose1996,oosterkamp2013}.

One common prerequisite of these experiments is the cooling of the oscillator to minimize thermal fluctuations and its coupling to environmental degrees of freedom.
This can be achieved either passively (as e.g. in Refs.~\cite{lahaye2004,cattiaux2021}), or by relying on active feedback cooling protocols typically borrowed from the optical domain~\cite{kleckner2006,poggio2007,teufel2011}.
Either way, low temperatures provide an advantage.

The thermodynamic cooling of such systems comes with long-standing challenges.
While the ground-state cooling of GHz oscillators can be achieved with a standard dry dilution refrigerator, temperatures below 1~mK are a requirement for observing quantum effects in the motion of oscillators in the MHz~\cite{cattiaux2021} or kHz range.
Conquering the sub-mK domain, which requires dedicated cryogenic techniques such as adiabatic nuclear demagnetization~\cite{pobell,abe2014}, is currently the subject of a collaborative research effort~\cite{pickett2018}.
A parallel challenge is the need for a vibration isolation system that can screen the oscillator from the omnipresent low-frequency mechanical vibrations afflicting any dilution refrigerator, wet or dry.

In this work, we report progress towards the simultaneous resolution of both challenges in a cryogenic setup dedicated to MRFM.
We show that it is possible to lower the temperature achievable in our setup by more than one order of magnitude -- from $T\approx$20 mK to $T\lesssim$1 mK -- without compromising the vibration isolation.
To achieve this, we combine the use of an adiabatic nuclear demagnetization stage with the mass-spring system for the filtering of mechanical vibration already described in Ref.~\cite{dewit2019}.
The lowest temperatures reached are established through the in-house implementation of a primary magnetic flux fluctuation thermometer (MFFT)~\cite{fleischmann2020}.
We observe an increase of the cantilever quality factor $Q$ upon lowering temperature, compatible with a power-law $Q^{-1}\propto T^{1/5}$.
Furthermore, we are able to detect its thermal motion down to 20 mK without observing saturation, indicating that we are limited by the sensitivity of the readout circuit rather than by uncontrolled heating of mechanical modes.

\section{Experimental setup}\label{sec:setup}

\begin{figure}[t]%
\centering
\includegraphics[width=\textwidth]{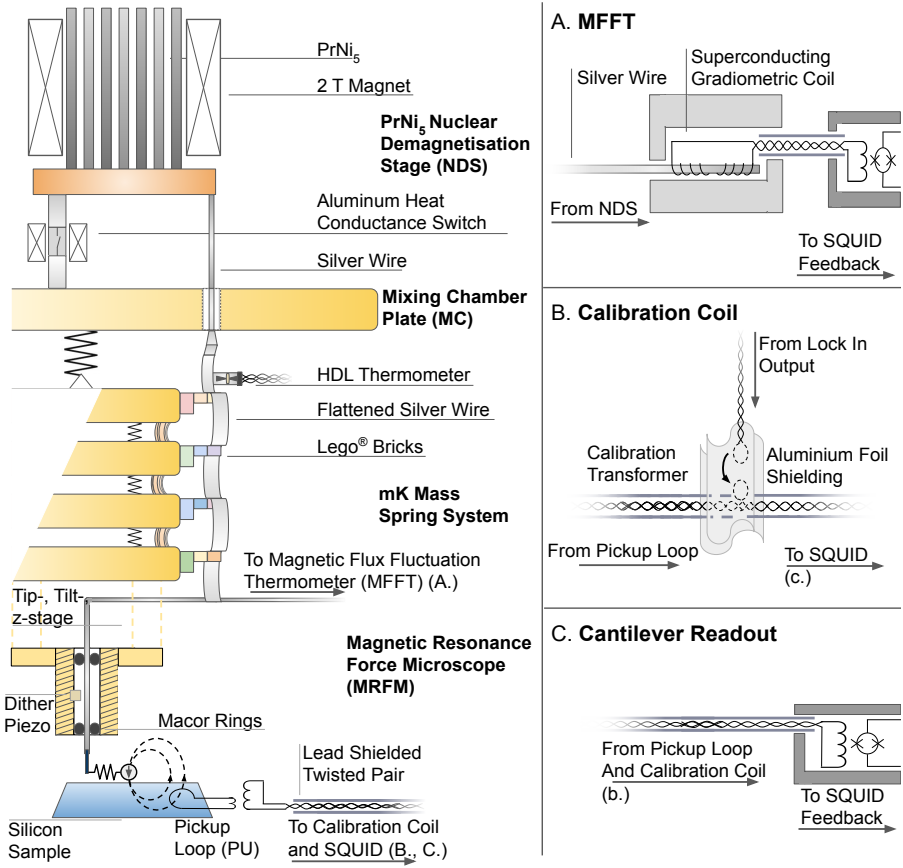}
\caption{Schematic drawing of the experimental setup. The main panel shows the adiabatic nuclear demagnetization stage, the mass-spring system for vibration isolation, and the MRFM system. The insets on the right show details on the magnetic flux fluctuation thermometer (MFFT) and the circuit connected to the MRFM sample, both of which are eventually readout via separate SQUIDs (Magnicon NC-1 Integrated Two Stage Current Sensor). An image of the setup is shown in Appendix~\ref{app:photo}.}\label{fig1:setup}
\end{figure}

Our experimental setup is described in Fig.~\ref{fig1:setup}.
At its core is the combination of an adiabatic nuclear demagnetization embedded within a pulse-tube dilution refrigerator~\cite{batey2013,todoshchenko2014,palma2017}.
We make use of a PrNi$_5$ nuclear stage for adiabatic nuclear demagnetization, positioned above the mixing chamber plate.
The magnetization of the PrNi$_5$ stage is controlled by a 2~T magnet, while the thermal contact with the mixing chamber can be turned on and off via an Aluminum heat conductance switch controlled by a separate magnet.
The 2~T magnet needs up to 40~A of current and is thermalized to the still plate, while the Aluminum switch needs up to 150~mA and is thermalized to the 50~mK plate.

Below the mixing chamber there is a mass-spring system optimized for passive filtering of mechanical vibrations between 100 Hz and 10 kHz.
It consists of four copper masses ($\approx$ 2 kg) connected by springs, that can easily be replaced to vary the resonance frequency. It is designed to filter both vertical and lateral vibrations, and it is thermalized to the mixing chamber plate.
We refer to Ref.~\cite{dewit2019} for more details on the mass-spring system and its performance in the dilution refrigerator before the installation of the demagnetization stage.

The magnetic resonance force microscope is firmly attached to the lowest mass of the mass-spring system.
The force sensor is an IBM-style silicon cantilever with a permanent Nd$_2$Fe$_{14}$B magnet attached to its tip.
The mass of the magnet is estimated to be $m\approx 1.5$~ng, which, together with the measurement of the frequency of the fundamental mode, $\omega_c\approx 653$~Hz, yields a stiffness $k\approx 2.6\cdot 10^{-5}$~N$\cdot$m$^{-1}$.
The tip-holder is connected to a plate that can be tip-tilted and lifted in the z-direction by three piezo knobs.
This allows to control the positioning of the cantilever with respect to the sample below it.
Oscillations of the cantilever can be electrically driven via a dither piezo.
The cantilever, the sample as well as their holders have been described in Ref.~\cite{dewit2019thesis}, to which we refer for further details.

A silver wire running down from the PrNi$_5$ provides the thermal link to cool down the cantilever holder.
After passing through the mixing chamber plate, the wire is anchored to the mass-spring system in order not to compromise the vibration isolation.
This is achieved by gluing the wire to LEGO\textsuperscript{\textregistered} blocks attached to each mass of the mass-spring system.
The extremely low thermal conductivity of LEGO\textsuperscript{\textregistered} blocks in the mK range~\cite{chawner2019} guarantees the thermal isolation of the silver wire from the mixing chamber without the need to compromise the vibration isolation.

The temperature of the silver wire is monitored through two different thermometers.
The first thermometer is a resistance thermometer developed by one of us (W.B.) and mounted on the silver wire right above the vibration isolation stage.
This reference thermometer offers fast, highly-accurate temperature readings between 15~mK and 1~K.
The second thermometer is a magnetic flux fluctuation thermometer (MFFT) built in-house, positioned below the mass-spring system and right above the cantilever holder.
The MFFT consists of a superconducting coil gradiometrically wound around the silver wire~\cite{palma2017}, see panel (a) in Fig.~\ref{fig1:setup}.
It is connected to a SQUID for the readout of the magnetic flux fluctuations induced in the coil by the thermal fluctuations of charge carriers in the silver wire~\cite{fleischmann2020}.

The three joints connecting the silver wire to the PrNi$_5$, the MFFT and the cantilever holder were spot-welded in-situ, silver-to-silver, rather than clamped.
This is done in order to maximize the thermal conductivity and avoid bottlenecks in the thermalization of the MFFT and the cantilever to the nuclear demagnetization stage.

Finally, the silicon sample below the MRFM sensor contains a pick-up loop for the magnetic readout of the cantilever position via a superconducting quantum interference device (SQUID)~\cite{usenko2011,dewit2019thesis}.
The pick-up loop is first connected by Aluminum wire bonds to a transformer chip situated in the sample holder.
The latter is in turn connected to a commercial SQUID via approximately 45 cm of superconducting twisted pair inside a lead capillary.
As shown in panels (b) and (c) of Fig.~\ref{fig1:setup}, the twisted pair cabling also includes a single-loop calibration transformer connected to a lock-in, that can be used to calibrate the transfer function of the readout circuit, as described in App.~\ref{app:calibration}.

\section{Results}\label{sec:results}

\subsection{Noise thermometry}

We begin the discussion of the experimental results from the thermometry, which is essential to establish that our setup is capable to reach temperatures below the typical base temperature, $\approx 20$~mK, of a dry dilution refrigerator.
In a noise thermometer such as the MFFT, the temperature is extracted from a measurement of the power density spectrum (PSD) of the noise registered by the device: in this case, the noise PSD $S_\Phi$ of the magnetic flux picked up by the coil of the MFFT.

\begin{figure}[t]%
\centering
\includegraphics[width=\textwidth]{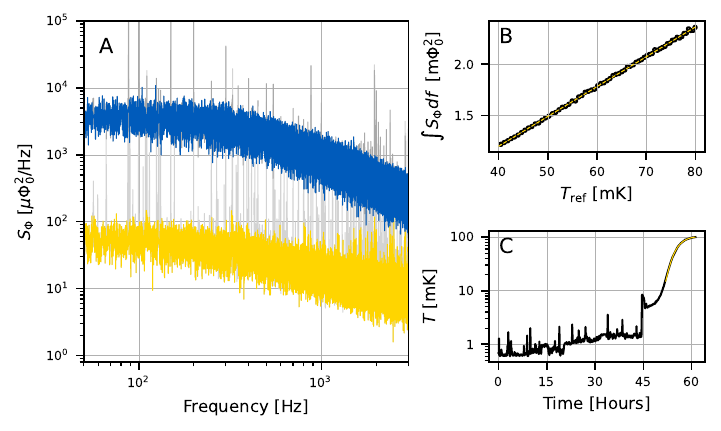}
\caption{\emph{Panel A}: Power spectral density of the magnetic flux noise measured in the magnetic flux fluctuation thermometer, at two different temperatures. \emph{Panel B}: Calibration of the noise thermometer versus the HDL resistance thermometer. The yellow dashed line is the linear fit used in the calibration of the thermometer. \emph{Panel C}: Temperature of the silver wire at the bottom of the vibration isolation stage, from the calibrated MFFT thermometer. The yellow curve shows the reference temperature in the range in which it is reliable.}\label{fig2:thermometry}
\end{figure}

Two typical spectra corresponding to different temperatures are shown in Fig.~\ref{fig2:thermometry}A.
They feature a plateau at low frequencies and a roll-off at high frequency, with a cutoff frequency at about 500 Hz.
In theory, the frequency dependence of the magnetic flux noise in a MFFT is expected to take the form~\cite{fleischmann2020}:
\begin{equation}\label{eq:mfft_theory}
S_\Phi = 4k_BT\,\sigma\,\mu^2_0 R^3\,G(R/\delta)
\end{equation}
where $\sigma$ is the bulk conductivity of the metallic wire which generates the magnetic fluctuations, $R$ its radius, $T$ its temperature, and $\mu_0$ its magnetic permeability.
The dimensionless function $G$ depends on the precise geometry of the MFFT (i.e., on the combined system composed of the metallic wire and the pick-up coil).
It encodes the frequency dependence of the noise via the skin depth $\delta(\omega)=\sqrt{2/\mu_0\sigma\omega}$ and it determines both the cut-off frequency and the roll-off of the spectral density with frequency (possibly together with additional effects contributed by the surrounding readout circuit, e.g. the cutoff introduced by amplifiers at the end of the measurement chain).

On top of the behavior predicted by Eq.~\eqref{eq:mfft_theory}, the measured spectra show many non-thermal interference peaks, presumably due to electrical interference picked up in the SQUID wiring, through crosstalk from the silver wire to the SQUID via the coil wound around the wire (see Fig.~\ref{fig1:setup}A), or because of interference further down the amplification chain. In order to extract the thermal noise signal, we remove the interference peaks during the data post-processing, using an empirical procedure described in App.~\ref{app:mfft}.
For consistency, the set of frequencies associated with interference peaks is determined once and kept common to all datasets.
In Fig.~\ref{fig1:setup}, the raw data is shown in grey below the post-processed data, explicitly showing the interference peaks removed in the data analysis.

Once the thermal noise signal is isolated, Eq.~\eqref{eq:mfft_theory} allows in principle to extract the temperature via a fit of the measured PSDs.
However, it is non-trivial to determine the function $G$ in Eq.~\eqref{eq:mfft_theory} from first principles, and furthermore we do not know precisely the conductivity $\sigma$ (or equivalently the residual resistivity ratio) for our silver wire.
Therefore, we resort to a calibration of our MFFT using the HDL reference thermometer.
Consistent with the expectation that the MFFT should operate as a primary thermometer, the spectral noise power below the cutoff frequency in the PSD varies linearly with the temperature, see Fig.~\ref{fig2:thermometry}B.
The unknown proportionality factor is determined empirically via a direct comparison to the reference thermometer temperature in the range between 40~mK and 80~mK, where the latter is reliable.
The linear behavior can then be extrapolated to lower temperatures, only limited by the noise floor of the SQUID readout circuit of the MFFT, which is approximately 1 $\mu\Phi_0/\sqrt{\text{Hz}}$.
More details on the MFFT calibration procedure are given in App.~\ref{app:mfft}.

In Fig.~\ref{fig2:thermometry}C we show the MFFT thermometry data during a 60-hour measurement run which started at the cold end of an adiabatic demagnetization cycle.
The data reveals that the temperature of the Ag wire at the bottom of the vibration isolation stage was as low as $0.6$ mK, very close to the theoretical lower limit reachable in our setup, which is set by the magnetic ordering of PrNi$_5$~\cite{pobell}.
We attribute the sharp temperature spikes visible during this time interval to the spurious presence of strong mechanical excitations of the system causing temporary heating of the silver wire. We find that these spikes often occur in the first week after the initial cool down from room temperature.
In the first part of the measurement run, in which both heaters and demagnetization stage were kept idle, the temperature slowly drifts upwards, approaching 2 mK after 45 hours.
At this point we applied heat via a heater, thermally anchored to the mixing chamber, causing the temperature to increase to 100 mK.
The aforementioned calibration was performed during this time window at the end of this nuclear demagnetization run.

\subsection{Temperature dependence of the quality factor}

Our second set of results concerns the dissipation properties of the cantilever.
Besides being a basic characterization of the MRFM system, we are motivated by the fact that applications of our experimental setup to CSL exclusions studies benefit from large quality factors of the cantilever force sensor.

\begin{figure}[t]%
\centering
\includegraphics[width=\textwidth]{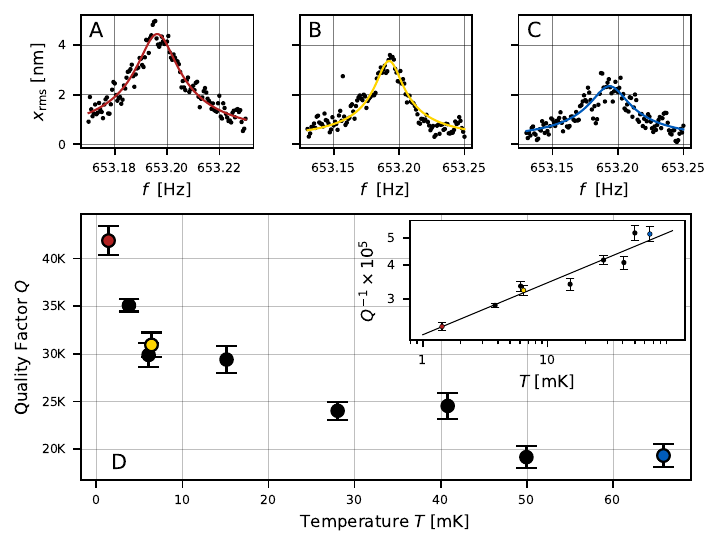}
\caption{\emph{Panel A-B-C}: Three examples of Lorentzian fits of the cantilever response to sweeps of the dither piezo driving frequency, at fixed driving amplitude and at three different temperatures, increasing from left to right. \emph{Panel D.} Temperature dependence of the quality factor extracted from the frequency sweeps. The red, yellow and blue data points correspond to panels A, B, C. The inset contains shows the sama data point against a fit to a power-law $Q^{-1}\propto T^{\alpha}$, which gives $\alpha = 0.19\pm 0.02$. See App.~\ref{app:qfactor} for more details on the data analysis.}\label{fig3:qfactor}
\end{figure}

The quality factor of the force sensor can be determined by driving it externally via the dither piezo and varying the frequency of the drive around the resonant frequency $\omega_c \approx 653$~Hz.
The displacement of the oscillator as a function of frequency can be determined from the voltage measured across the readout SQUID, using the calibration described in App.~\ref{app:calibration}.
We performed multiple frequency sweeps for different temperatures and fixed driving amplitudes, of which some examples are shown in Fig.~\ref{fig3:qfactor}A-C.
They display typical resonant behavior: a Lorentzian peak with maximum amplitude $A$ of a few nanometers and width $\Gamma$ on the order of a few tens of milliHertz.
The direct comparison of panels A to C in Fig.~\ref{fig3:qfactor} indicates that the peak broadens at higher temperatures, indicating an increase in dissipation.

The quality factor $Q = \omega_c/\Gamma$
can be extracted via a fit of each frequency sweep to a Lorentzian peak.
The results are shown in Fig.~\ref{fig3:qfactor}D and they reveal that the quality factor slowly increases with decreasing temperature: from $Q\approx 2\cdot 10^4$ at $T\approx 70$~mK to $Q\approx 4\cdot 10^4$ at $T\approx 1.5$~mK.
We refer to App.~\ref{app:qfactor} for details of the data analysis and the error estimation.

We note that for these measurements, the cantilever was positioned approximately 100~$\mu$m away from the sample in order to minimize the influence of surface two-level systems, presumably dangling bonds that cause surface spins, on the cantilever's properties~\cite{denhaan2015,dewit2018}.
Thus, the increase in $Q$ must likely be connected to a factor intrinsic to the cantilever.
A fit to a power-law $Q^{-1}\propto T^{\alpha}$, see inset of Fig.~\ref{fig3:qfactor}D, yields an exponent close to $\alpha=0.2$.
Similar scaling has been previously observed in different types of micro-resonators~\cite{hutchinson2004,imboden2014}, and may be indicative of the presence of structural disorder in the silicon cantilever.
Future measurements with an improved coupling should allow to improve the accuracy of this temperature dependence.

\subsection{Thermal motion of the force sensor}

The energy stored in the oscillator in the absence of any external drive, including that originating from unwanted vibrations of the setup, is subject to thermal fluctuations.
Therefore, to demonstrate that the silver wire does not impair the performance of the vibration isolation stage, we set out to measure the thermal motion of the cantilever versus temperature.

The fluctuations of the cantilever displacement can be inferred from a measurement of the PSD of the MRFM readout circuit, where the thermal motion appears as a Lorentzian peak centered around $\omega_c$.
The area under the peak can be converted into an rms displacement of the cantilever, $\langle x^2 \rangle$, where $\langle \cdot \rangle$ denotes the time-average of the displacement over the duration of each measurement.
Since the measurement time ($t_\textrm{meas} = 600$~s) is much longer than the ring-down time of the cantilever ($\tau=Q / \pi\omega_c \approx 5$~s for $Q=10^4$), the rms displacement can be converted to temperature via the equipartition theorem, $k_BT_\textrm{cantilever} = k\langle x^2 \rangle$, where $k$ is the stiffness of the cantilever.
In order for this measurement to be successful, it is important to have a large sensitivity of the readout circuit to the oscillator displacement.
Therefore, to measure the thermal motion, we moved the cantilever to be much closer to the sample with respect to the position used in the previous Section.
Due to the influence of two-level systems on the sample surface, the quality factor through this measurement was lower than in the previous one: it did not exceed $2\cdot 10^4$ and decreased with temperature~\cite{denhaan2015,dewit2018}.

\begin{figure}[t]%
\centering
\includegraphics[width=\textwidth]{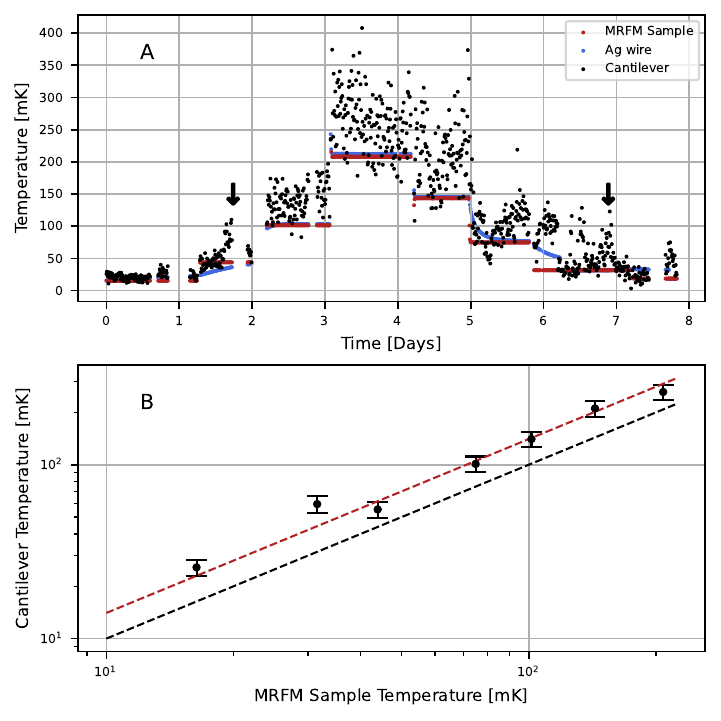}
\caption{\emph{Panel A}: Cantilever temperature, MRFM sample temperature and silver wire temperature measured versus time. The cantilever temperature is estimated from the displacement fluctuations of the cantilever. Time zero on the horizontal axis correspond to 16:03 on Sunday, January 3rd 2022.
Black arrows indicate time intervals likely corresponding to a measurable increase in vibrations, e.g. due to activity in the building.
\emph{Panel B}: Averaged cantilever temperature from panel (A) plotted against the MRFM sample temperature. The black dashed line indicates the condition $T_\textrm{cantilever}=T_\textrm{sample}$. The red dashed line is a fit $T_\textrm{cantilever}=c T_\textrm{sample}$, yielding a coefficient $c=1.40\pm 0.07$.}\label{fig4:thermalmotion}
\end{figure}

In Fig.~\ref{fig4:thermalmotion}A we show the results accumulated over the course of a week-long measurement run.
Over this time, we varied the temperature of the MRFM sample containing the readout circuit in a step-like fashion, from approximately 15~mK up to approximately 200~mK and back; the sample temperature was measured via a second reference thermometer (not shown in Fig.~1).
All the while, we monitored the rms fluctuations of the cantilever motion, via a PSD measurement with a 50 kHZ sampling rate and a measurement time of ten minutes; see App.~\ref{app:thermalmotion} for further details.
The results, cast in terms of a cantilever temperature, are shown in  Fig.~\ref{fig4:thermalmotion}, together with the temperature $T_\textrm{bath}$ of the tip holder of the cantilever chip.
We find that the variations of the oscillator temperature largely follows the bath temperature, up to fluctuations.
In Fig.~\ref{fig4:thermalmotion}A we also show the temperature of the reference thermometer anchored to the silver wire (the same used in Fig.~\ref{fig2:thermometry}), which relaxes to that of the MRFM sample during each step.
The heat path from the sample to the silver wire may go through the cantilever itself or through the Macor rings (see Fig.~\ref{fig1:setup}).

Averaging over measurements of the oscillator temperature obtained at constant bath temperatures yields the bottom plot of Fig.~\ref{fig4:thermalmotion}.
We see that although $T_\textrm{cantilever} > T_\textrm{sample}$, there are no clear signs of saturation of the cantilever temperature down to the limit of our signal-to-noise sensitivity at $\approx$20 mK.
Note that the error bars in Fig.~\ref{fig4:thermalmotion}B include both the statistical uncertainty coming from the averaging (i.e., the standard error of the mean) and the calibration uncertainty discussed in App.~\ref{app:calibration}, with the latter providing the most important contribution to the total uncertainty.
In fact, we cannot exclude that the cantilever temperature is overestimated due to a systematic error in the calibration procedure of App.~\ref{app:calibration}.
This possibility is flagged by a linear fit to the data in Fig.~\ref{fig4:thermalmotion}B, assuming a zero intercept, which yields a linear coefficient incompatible with one -- compare the red and black dashed lines.

At fixed $T_\textrm{sample}$, subsequent measurements of $T_\textrm{cantilever}$ are expected to fluctuate around the average value $\bar{T}_\textrm{cantilever}$ with a standard deviation $\delta T~ =~\sqrt{\tau/t_\textrm{meas}}\,\bar{T}_\textrm{cantilever}$.
Therefore, for each plateau in $T_\textrm{sample}$ in Fig.~\ref{fig4:thermalmotion}A, approximately 95\% of the data points are expected to fall within an interval $4\,\delta T$ around the average cantilever temperature.
For $Q=10^4$, the confidence interval is approximately 37\% of $\bar{T}_\textrm{cantilever}$.
While this is indeed the case for much of the time, during some time intervals we observe systematic fluctuations outside of this confidence interval, see e.g. the two black arrows in Fig.~\ref{fig4:thermalmotion}A.  
This could hint at an additional, possibly co-existing, origin of the temperature difference between the cantilever and the sample.
Namely, it could indicate that the cantilever still suffers from vibrations of the building to a measurable degree.

\section{Conclusions}\label{sec:conclusions}

We demonstrated that it is possible to combine nuclear demagnetization and vibration-isolation techniques to reach a vibrationally quiet 1 mK environment at the bottom of a dry dilution refrigerator.
Improvements to the SQUID circuit should allow us to improve the sensitivity of the readout by as much as one order of magnitude.
In turn, this may allow us to probe the thermal motion of the cantilever from the 20 mK demonstrated here to the sub-mK regime, far away from the surface with the pick-up coil.
Even lower temperatures may be reached by adding a second, copper-based nuclear demagnetization stage thermalized to the silver wire at the bottom of the vibration-isolation stage.

Looking ahead, we can discuss the feasibility of our setup for improved tests of CSL models~\cite{carlesso2022}.
To this end, two important experimental parameters are the thermal force noise $S_F$ and the radius $R$ of the magnet attached to the cantilever.
Getting a lower force noise with a larger diameter particle will improve the exclusion plot for CSL parameters such as the collapse rate $\lambda$.
Previous results from our group~\cite{usenko2011,vinante2016} reached $\sqrt{S_F}\approx5\cdot 10^{-19}\,\textrm{N}/\sqrt{\text{Hz}}$ with a magnet radius of $R\approx2.3~\mu$m, smaller than our present value of $3.6~\mu$m.
Extrapolating our current results, we may estimate the thermal force noise of our cantilever~\cite{stowe1997} assuming that thermal motion can be detected down to $T\approx 0.5$~mK, at which we can estimate $Q\approx 50\cdot10^3$ based on the trend shown in Fig.~\ref{fig3:qfactor}D.
This would allow us to reach
\begin{equation}
\sqrt{S_F} = \sqrt{\frac{4k_BT\,k}{\omega_c Q}}\approx 6\cdot 10^{-20}\,\textrm{N}/\sqrt{\text{Hz}}\,,
\end{equation}
As the figure of merit for CSL measurements is roughly $S_F/R^2$, the comparison of past and current values of $S_F$ and $R$ suggests a possible improvement of two orders of magnitude in the exclusion plot for $\lambda$.

The observation of the thermal motion of the cantilever down to temperatures $T\approx 20$~mK implies that, at least during quiet times, the vibration isolation is good enough to attenuate the cantilever position noise at resonance to a level below $\sqrt{4 k_BT / k}\approx 0.2$~nm. Note that with a quality factor $Q\approx 4\cdot 10^4$, this requires the cantilever base to vibrate by less than $50$~fm in the cantilever bandwidth. The position noise will have to be improved by a factor of approximately five in order to observe thermal motion at 1~mK.
In order for a future measurement to put the tightest constraints on CSL parameters, we need to also work on an improved accuracy of the parameter that converts the SQUID voltage to the motion of the resonator, which will require a careful characterisation of the SQUID and the readout circuit itself.

\backmatter

\bmhead{Data availability statement}

The data and code that have been used to generate the results of this work are available on \href{https://doi.org/10.5281/zenodo.6897977}{Zenodo}.

\bmhead{Acknowledgments}

We thank K. Heeck, A. Vinante, H. Ulbricht, and G. Pickett for useful discussions. 
This publication is part of the projects 175.010.2015.050, 680-91-123 and 2019.088 financed by the Dutch Research Council (NWO).

\clearpage
\begin{appendices}

\section{Image of the experimental setup}
\label{app:photo}

 \begin{figure}[h]
\centering
\includegraphics[width=1\textwidth]{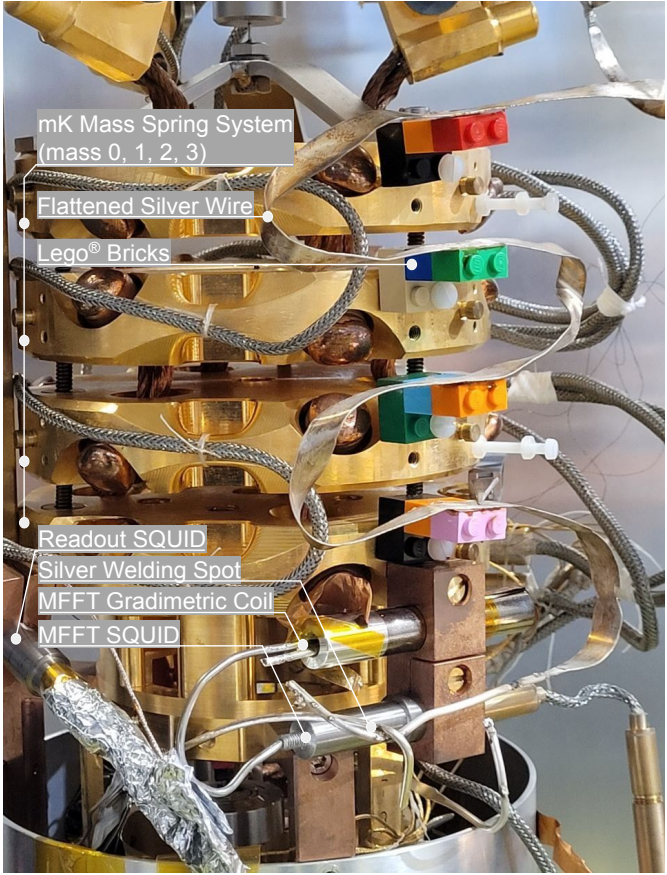}
\vspace*{8mm}
\caption{Photo of the the experimental apparatus of Fig.~\ref{fig1:setup}. The mass-spring system and the Ag wire with the LEGO\textsuperscript{\textregistered} blocks are visible, as well as the the lead casing of the readout SQUID, the MFFT and the MFFT SQUID. Note that this picture was taken before some of the final spot welds were made to connect the different segments of the Ag wire.}\label{fig_supp:readout_circuit}
\end{figure}

\clearpage
\section{Magnetic flux fluctuation thermometry: data analysis}
\label{app:mfft}

The results reported in  Fig.~\ref{fig2:thermometry} are obtained from a continuous monitoring of the magnetic flux fluctuation thermometer (MFFT) over the course of about 60 hours, between December 21st and December 24th, 2021. This measurement run started at the end of an adiabatic demagnetization cycle, thus at the lowest temperature achieved in the Ag wire at the bottom of the vibration isolation stage. After 45 hours of idle monitoring, the temperature was gradually increased to approximately 100 mK over the course of 15 hours. This was done via a heater thermally attached to the mixing chamber plate, which caused heat to leak into the part of the system attached to the PrNi5 demagnetization stage.

Throughout this time, the signal from the MFFT-dedicated Magnicon SQUID was sampled at 2 MHz and stored on disk. About 13000 time traces were saved during the measurement run, each lasting approximately 16 s. 
We determined the power spectral density (PSD) of each time trace via a discrete Fourier transform. We used Welch's method with a Hamming window and 50\% overlap between consecutive windows. In order to improve the signal-to-noise ratio, we averaged the PSD obtained from ten consecutive time traces. Note that this yields a final time resolution of about 160 s for the determination of the temperature via the MFFT. The 1321 averaged spectra resulting from this procedure, with frequency resolution of 0.0625 Hz and an upper cutoff of 10 kHz, are made available as the raw data underlying Fig.~\ref{fig2:thermometry} (the full data is available upon request).

As mentioned in the main text, on top of the behavior expected from Eq.~\eqref{eq:mfft_theory}, the measured spectra show many non-thermal interference peaks which need to be removed in post-processing for a reliable calibration of the thermometer. We do so via the following procedure. For each measured spectrum, we:
\begin{enumerate}
    \item discard data below 50 Hz, which is affected by $1/f$ noise as well as noise originating from the pulse tube.
    \item divide the remaining data into smaller frequency bins of 50 Hz width.
    \item for each bin, determine the frequencies at which interference peaks occur by identifying outliers in the bin. The outliers are identified using the combination of two criteria: a height threshold, and a $z$-score threshold.
\end{enumerate}
If a certain frequency is identified as an outlier in more than 2.5\% of the measured PSD, it is discarded from all the spectra for the remaining of the post-processing analysis.
Spectra measured at high temperatures are not used in determining this threshold since the interference peaks are drowned out by the signal.
In Fig.~\ref{fig2:thermometry}, the raw spectra are plotted in grey, while the cleaned spectra in yellow and blue.
Discarded data points are therefore visible in grey in the background of the post-processed spectra, revealing the effectiveness of the peak-removing procedure.

The calibration procedure of the MMFT is done as follows.
First, we numerically integrate the PSD to determine the flux noise power between a lower cutoff $\omega_0=50$ Hz and an upper cutoff $\omega_1=3$ kHz.
Consistent with the expectation that the MFFT should operate as a primary thermometer, the resulting spectral noise power,
\begin{equation}
P = \int_{\omega_0}^{\omega_1} S_\Phi \,d\omega\,,
\end{equation}
depends linearly on the temperature $T$ measured by the reference thermometer above 15 mK, as shown in the top right panel of Fig.~\ref{fig2:thermometry}.
A linear fit of the data in this temperature range provides the required calibration for the MFFT thermometer.
We used only the temperature range between 40 mK and 80 mK to perform the linear fit.
For this procedure to be valid, we must assume that in this temperature range the temperature along the silver wire was uniform across the vibration isolation stage, so that the MFFT and the reference thermometer are at the same temperature.
This calibration is subsequently applied to the entire measurement run. 

\section{Analysis and calibration of the two-stage cantilever readout circuit}
\label{app:calibration}

\begin{figure}[h]%
\centering
\includegraphics[width=\textwidth]{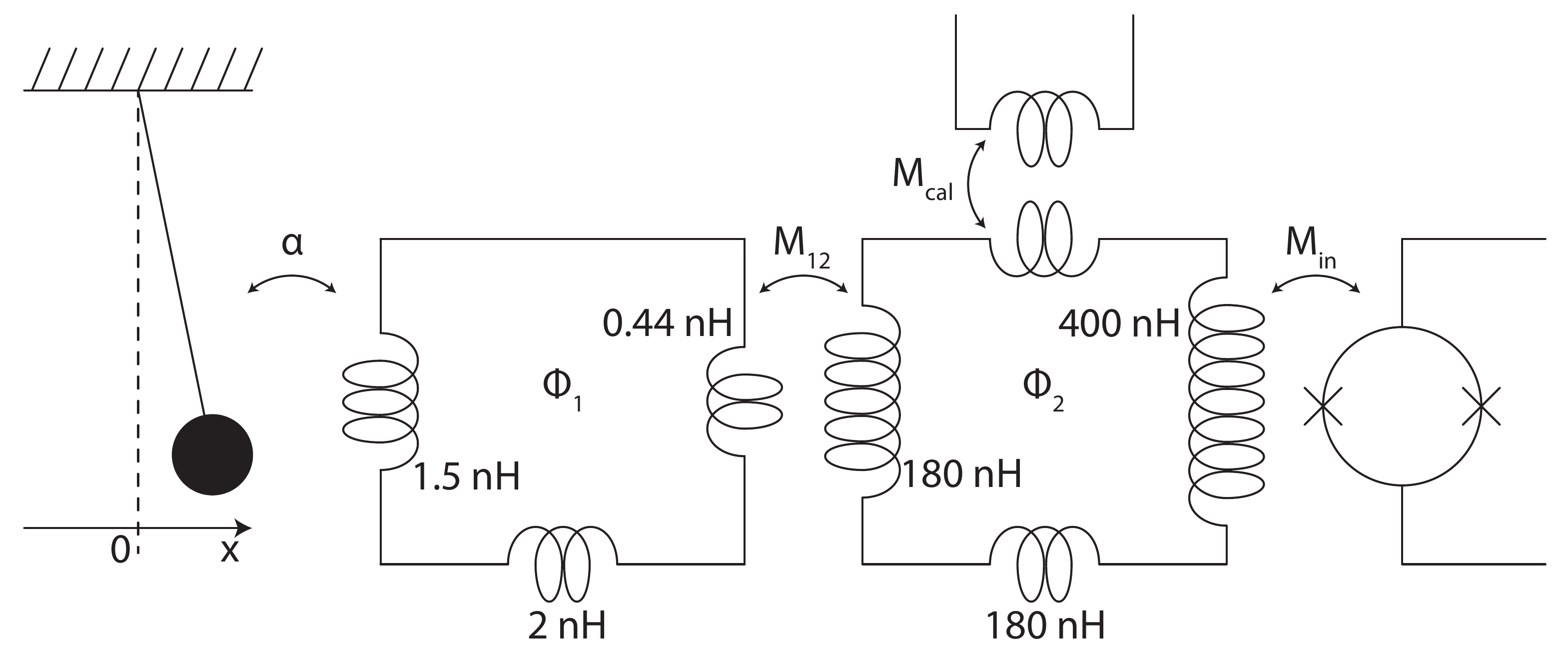}
\caption{Schematic circuit of the two-stage readout circuit for the cantilever force sensor. Here $\alpha$ denotes the linear coupling between the pick-up loop and the cantilever displacement, while the mutual inductance $M_{12}=5$~nH.}\label{fig_supp:readout_circuit}
\end{figure}

In Fig.~\ref{fig_supp:readout_circuit} we show a schematic circuit for the two-stage readout circuit for the cantilever force sensor.
The displacement $x$ of the oscillator from its rest position induces a change in the flux through the pick-up loop (with total flux $\Phi_1$) on the sample chip. For small $x$, the induced flux is linear in the displacement: $\Phi_\textrm{ind}=\alpha x$, where $\alpha$ is a coupling constant (with dimensions of Wb$\cdot$m$^{-1}$) to be determined: it depends on the position of the cantilever with respect to the pick-up loop, and in particular on the height of the cantilever with respect to the pick-up loop on the sample.

This change in flux induced by the cantilever motion is transferred first via a transformer to a secondary loop (with total flux $\Phi_2$) which contains the input coil of the readout SQUID.
The secondary loop also includes a calibration coil that can be used to externally inject a flux in the circuit.

\begin{figure}[t!]%
\centering
\includegraphics[width=\textwidth]{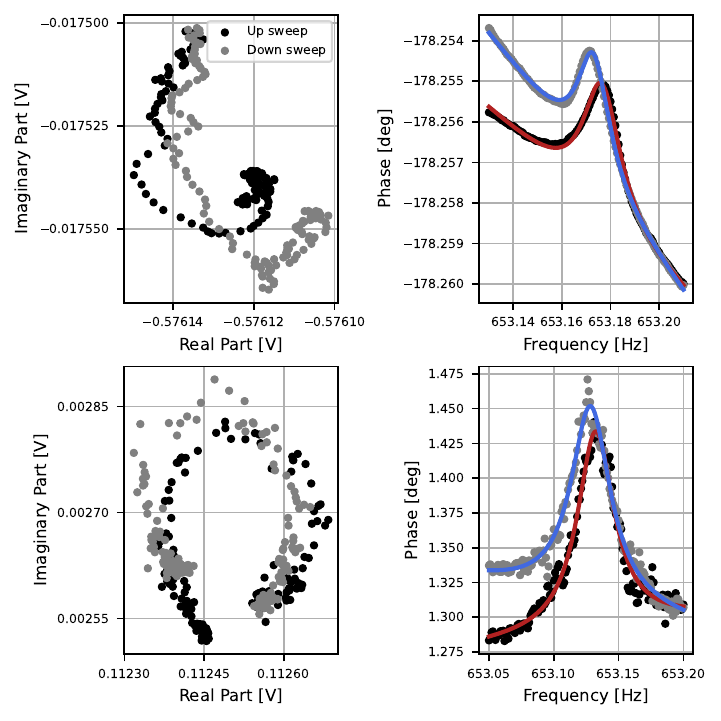}
\caption{Calibration fits to extract the coupling $\beta$. \emph{Top.} Calibration for the piezo sweep measurements used in Fig~\ref{fig3:qfactor}. In this case we extract $\beta = (3.87\pm0.03)\cdot 10^{-5}$. \emph{Bottom.} Calibration for the thermal motion measurements shown in Fig.~\ref{fig4:thermalmotion}. In this case we extract $\beta = (3.67\pm0.04)\cdot 10^{-4}$. The fits are to the peak function of Eq.~\eqref{eq:tan_phi_model}, plus a linear background term, as described in the text.}\label{fig_supp:calibration}
\end{figure}

Neglecting the back-action of the SQUID on the readout circuit, the potential energy of the latter of Fig.~\ref{fig_supp:readout_circuit} can be written as:
\begin{equation}
U = \frac{1}{2L_1}\Phi_1^2 + \frac{1}{2L_2}\Phi_2^2 + \frac{M_{12}}{L_1L_2} \Phi_1\Phi_2 + \frac{1}{L_1}\Phi_1\,\alpha x + \frac{1}{2}k_0x^2 - \frac{1}{L_2}\,\Phi_2\,\Phi_\textrm{cal}
\end{equation}
Here, $L_1$ and $L_2$ are the total inductances of the two stages of the circuit, $k_0$ is the bare spring constant of the oscillator, and $\Phi_\textrm{cal}=M_\textrm{cal} I_\textrm{cal}$ is the flux induced by the calibration coil. The stationary condition $\partial U /\partial\Phi_1=0$ yields the conservation law for the flux $\Phi_1$:
\begin{equation}
\Phi_1 = -\frac{M_{12}}{L_2}\Phi_2 - \alpha x\,.
\end{equation}
This relation can be inserted back in the potential to eliminate $\Phi_1$. The result is:
\begin{equation}
U = \frac{\eta}{2L_2}\Phi_2^2+\frac{1}{2}k x^2 - \frac{\alpha M_{12}}{L_1L_2}\,\Phi_2\,x - \frac{\Phi_\textrm{cal}\Phi_2}{L_2}\,,
\end{equation}
with $k=k_0-\alpha^2/L_1$ a modified spring constant due to the presence of the pick-up loop, and
\begin{equation}
\eta = 1 - \frac{M_{12}^2}{L_1 L_2}\,.
\end{equation}
In this form, the potential energy can be used to derive the equation of motion for the cantilever displacement $x$. Neglecting, for the moment, thermal noise as well as additional driving sources such as the piezo, one obtains (in frequency domain, denoted by hats):
\begin{equation}
\hat{x} =  \frac{\alpha M_{12}}{m L_1 L_2}\,\frac{1}{\omega_0^2-\omega^2+i\gamma \omega}\,\hat{\Phi}_2
\end{equation}
Here, $\gamma$ is the damping rate, $m$ is the mass of the cantilever, and $\omega_0=\sqrt{k/m}$ the resonant frequency. The flux $\Phi_2$, in turn, is the sum of the contributions induced by the calibration coil and by the oscillator displacement.
From the condition $\partial U/\partial \Phi_1=0$, we find:
\begin{equation}
\eta \Phi_2 = \Phi_\textrm{cal} + \frac{\alpha M_{12}}{L_1}\,x
\end{equation}
Hence, in frequency domain:
\begin{equation}
\eta \hat{\Phi}_2 = \hat{\Phi}_\textrm{cal} + \frac{\alpha^2}{m L_1}\,\frac{M_{12}^2}{L_1L_2}\,\frac{1}{\omega_0^2-\omega^2+i\gamma \omega}\,\hat{\Phi}_2
\end{equation}
With some algebra, we can obtain the response of the readout circuit to the calibration signal:
\begin{equation}
\hat{\Phi}_2 = \frac{1}{\eta}\,\frac{\omega_0^2-\omega^2+i\gamma\omega}{\omega_0^2-\delta^2-\omega^2+i\gamma\omega}\,\hat{\Phi}_\textrm{cal}\equiv T(\omega)\,\hat{\Phi}_\textrm{cal},
\end{equation}
where
\begin{equation}
\delta^2 = \frac{\alpha^2}{m L_1}\frac{1-\eta}{\eta}\,.
\end{equation}
The phase $\phi$ of complex-valued transfer function $T(\omega)$ is given by:
\begin{equation}
\tan \phi = \frac{\delta^2\,\gamma\,\omega}{(\omega_0^2-\omega^2)(\omega_0^2-\delta^2-\omega^2) + \gamma^2\omega^2}\,.
\end{equation}
Qualitatively, in the weak-coupling regime $\delta \ll \gamma \ll \omega_0$ valid for our experiment, the phase response shows a peak at $\omega \approx \omega_0$ of amplitude $\delta^2/\gamma\omega_0$ and width controlled by $\gamma$.
It is convenient to introduce the quality factor of the cantilever, $Q=\omega_0/\gamma$, as well as a dimensionless parameter,
\begin{equation}
\beta = \delta / \omega_0\,,
\end{equation}
which is a figure of merit of the sensitivity of the readout circuit to the cantilever displacement. Then, the phase response can be written as:
\begin{equation}\label{eq:tan_phi_model}
\tan \phi = \frac{Q\beta^2\,(\omega/\omega_0)}{Q^2(1-\omega^2/\omega_0^2)(1-\beta^2-\omega^2/\omega_0^2) + (\omega/\omega_0)^2}\,.
\end{equation}
This expression can be used to estimate the parameters $\beta, Q, \omega_0$ for the experimental setup.
This is done by applying a calibration signal $\Phi_\textrm{cal}$ of constant amplitude and varying frequency $\omega$, and then fitting the resulting phase response to Eq.~\eqref{eq:tan_phi_model}.
In Fig.~\ref{fig_supp:calibration} we show the results of two calibration sweeps essential for the results shown in the main text.
Note that, in practice, the data shows a linear drift in the phase which we attributed to a drift in the phase shift of the drive current.
This drift is associated with the capacitance of the wiring and a drift in the value of the resistor which converts the output voltage of the lock-in to a current injected in the calibration transformer; this linear drift is included as an additional free parameter in the fit.

In the top row of Fig.~\ref{fig_supp:calibration}, we show a calibration performed at the cantilever position used to study the temperature dependence of the quality factor (Fig.~\ref{fig3:qfactor} of the main text).
The resonator is positioned far from the surface with the pick-up loop to avoid interaction with the surface spins near the pick-up loop.
Due to the expected small coupling to the cantilever, the input data for this calibration measurement consists of the average of 24 frequency sweeps.
We note that while the absolute value of the response suffers from slow drift of the gain of the amplifiers used, even after the averaging, the phase response only suffers from the noise in the SQUID.
A non-linear fit of Eq.~\eqref{eq:tan_phi_model} yields $\beta = (3.77 \pm 0.02)\cdot 10^{-5}$, $Q = 37000 \pm 500$ and $\omega_0 = 653.17$ Hz.
The uncertainty on fit parameters was determined via a simple bootstrap procedure: the fit was performed 1000 times on resampled data, and the uncertainty obtained from the variance of the resulting distributions of fit parameters.

In the bottom row of Fig.~\ref{fig_supp:calibration}, we show the results for a calibration sweep taken at a much lower position of the cantilever.
In this case, due to the stronger coupling, a single sweep was enough to achieve a sufficient signal-to-noise ratio.
In this case, the fit yields $\beta = (3.69 \pm 0.03)\cdot 10^{-4}$, $Q = 18500 \pm 400$ and $\omega_0 = 2\pi \times 653.11$ Hz.
This value of the coupling strength $\beta$, ten times larger than in the previous position of the cantilever, was used in the data analysis of the thermal motion of the cantilever (Fig.~\ref{fig4:thermalmotion} of the main text).
Note that, comparing the two calibration measurements, we note a $\pi$ phase shift in the response of the cantilever, which we attribute to the different positioning of the latter with respect to the pick-up loop.

When the calibration coil is not used, the flux through the readout SQUID is directly proportional to the cantilever displacement $x$:
\begin{equation}\label{eq:x_to_flux}
\Phi_\textrm{SQUID} = \frac{M_\textrm{in}}{L_2}\Phi_2 = M_\textrm{in}\,\sqrt{\frac{k}{\eta L_2}} \,\beta\,x
\end{equation}
Thus, in order to determine the displacement $x$ from a measurement of $\Phi_\textrm{SQUID}$, we need to know the spring constant $k$ as well as the circuit parameter $\eta L_2$.
The former can be estimated from the frequency of the cantilever and the mass $m$ of the Nd$_2$Fe$_14$B magnet on its tip.
From a measurement of the radius of the magnet, $R\approx 3.65\cdot 10^{-6}$~m, assuming that the sphere outweighs the silicon cantilever and assuming a spherical shape for simplicity, as well as knowing the mass density $\rho=7600$~kg$\cdot$m$^{-3}$, the mass of the oscillator turns out to be
\begin{equation}
m \approx \frac{4\pi R^3}{3}\,\rho \approx 1.54\cdot10^{-9}\,\textrm{g}.
\end{equation}
Hence we obtain:
\begin{equation}
    k \approx \,m\omega_0^2 \approx 2.6\cdot10^{-5}\,\text{N}\!\cdot\!\text{m}^{-1}\,.
\end{equation}

The circuit parameter $\eta L_2$ is estimated as follows.
The transformer connecting the two loops has a gradiometric washer design with a total mutual inductance of
\begin{equation}
M_{12}\approx5~\textrm{nH}\,,
\end{equation}
Additionally, the transformer has an input inductance of $\approx 2\cdot90$~nH on the SQUID side and of $2\cdot0.22$~nH on the pick-up loop side.
The inductance of the secondary loop is the sum of contributions coming from: (1) the input inductance of the SQUID, 400~nH; (2) the inductance of $\approx$~45 cm of twisted pair, which we estimate to be $\approx$~180 nH. Hence we have:
\begin{equation}
L_2 \approx (400 + 180 + 2\cdot90)~\textrm{nH} \approx 760~\textrm{nH}
\end{equation}
The inductance of the primary loop is the sum of the pick-up loop inductance on the MRFM sample, which is $\approx 1.5$~nH, the contribution coming from the transformer and the inductance from the loop formed by the bonding pads and bonding wires to the transformer chip. The latter contributions we estimate to be $\approx 2$~nH based on the bonding area. Hence
\begin{equation}
L_1\approx (1.5 + 2 + 2\cdot 0.22)~\textrm{nH}\approx 4~\textrm{nH}
\end{equation}
These inductances yield
\begin{equation}
\eta \approx 0.99\,.
\end{equation}
Finally, the input coupling of the SQUID was $M_\textrm{in}=2\cdot10^6~\Phi_0\cdot$A$^{-1}$.
Plugging in these numbers in Eq.~\eqref{eq:x_to_flux}, we obtain a conversion factor:
\begin{equation}\label{eq:calibration_coefficient}
c \equiv M_{\textrm{in}}\sqrt{\frac{k}{\eta L_2}} \approx 0.012\; \Phi_0\cdot\textrm{nm}^{-1}\,.
\end{equation}
This factor can be used to obtain the displacement given a measurement of the flux through the SQUID.
In practice, what we measure is the output voltage $V_\textrm{output}$ (in Volts) of the SQUID operated in FLL mode, with a transfer function of $0.43~\textrm{V}/\Phi_0$ and a gain $G=10$ due to the SRS amplifier at the end of the measurement circuit.
The conversion back to the oscillator displacement is done as follows:
\begin{equation}\label{eq:calibration_V_to_x}
x = \frac{V_\textrm{output}}{0.43 \cdot G \cdot c \cdot \beta}\,\textrm{nm}\,.
\end{equation}
It is hard to estimate the individual uncertainty associated with all the circuit parameters entering Eq.~\eqref{eq:calibration_coefficient}.
Thus, to estimate the calibration uncertainty associated with Eq.~\ref{eq:calibration_V_to_x} we assign a relative uncertainty $\delta c / c = 0.1$.
To compute the uncertainty on $x$, this uncertainty can then be combined with the uncertainty on $\beta$ coming from the calibration fit of Fig.~\ref{fig_supp:calibration} as well as, if appropriate, the statistical uncertainty associated with the output voltage when averaging over many measurements.
Note that this uncertainty may not reflect the presence of an unknown systematic error in the calibration stemming from, e.g., a wrong estimate of the loop inductances.

\section{Temperature dependence of the quality factor: data analysis}
\label{app:qfactor}

The measurements reported in Fig.~\ref{fig3:qfactor} took place during a demagnetization sweep  which ran between Dec.~11 and Dec.~13, 2021.
The magnet current was ramped down in a step-like fashion from 40~A down to zero.
At each step, we measured the response of the cantilever to a drive applied to the piezo, varying the driving amplitude and sweeping the driving frequency.
The cantilever was read out using the two-stage SQUID detection circuit described in Appendix~\ref{app:calibration}, with the SQUID signal converted to an estimate of the displacement of the oscillator induced by the piezo drive.
We only report results corresponding to the largest drive amplitude, i.e. 10~mV applied to the input of the piezo.

The temperature of the Ag wire throughout the demagnetization sweep was monitored using the MFFT, and determined in post-processing using the same calibration described in Appendix~\ref{app:mfft}.
We assume that the cantilever-piezo system was in thermal equilibrium with the Ag wire throughout the measurements.
We assign a temperature to each step of the demagnetization ramp by averaging out the MFFT measurements that occurred during each step; these temperature values are then assigned to each frequency sweep depending on the initial time of the sweep.

The frequency sweeps display the typical resonant behavior around a center frequency $\omega_c\approx 653$ Hz.
We swept the frequency from a value below $\omega_c$ to a value above it and back down.
Because of the presence of hysteresis between the ``up'' and ``down'' portions of each sweep, they were analyzed separately.
The slight hysteresis is caused by the slow response of the resonator because of its high Q-factor,
From each displacement vs frequency curve, we obtain the quality factor through a fit to a Lorentzian peak function.
The uncertainty on the quality factor was determined via a simple bootstrap procedure: we re-sampled the data with replacement 1000 times, performed the non-linear fit on each re-sampled curve, and assigned uncertainty based on the empirical distribution of fit parameters thus obtained.
Results from the ``up'' and ``down'' sweeps, as well as from different sweeps that occurred at the same temperature, were aggregated using inverse-variance weighting to obtain the quality factor vs temperature data shown in Fig.~\ref{fig3:qfactor}.

\section{Detection of the thermal motion of the cantilever: data analysis}\label{app:thermalmotion}

The measurement run summarized in Figure 4 took place over the course of almost 200 hours between January 4th and January 11th, 2022.
With minor interruptions, the voltage across the SQUID coupled to the cantilever detection signal was sampled at 50~kHZ.
In total, 950 time traces each lasting approximately 10 minutes were stored on disk during this time interval.
For each time trace we computed the power spectral density with 1~mHz resolution. 
The power spectral density was obtained via Welch's method, as described in Sec.~\ref{app:mfft} for the MFFT data.
After a supervised visual inspection of the data, 28 time traces had to be discarded because they displayed clear artifacts related to the feedback loop of the SQUID going out of lock or other miscellaneous anomalies preventing the continuation of the data processing.

\begin{figure}[t!]%
\centering
\includegraphics[width=\textwidth]{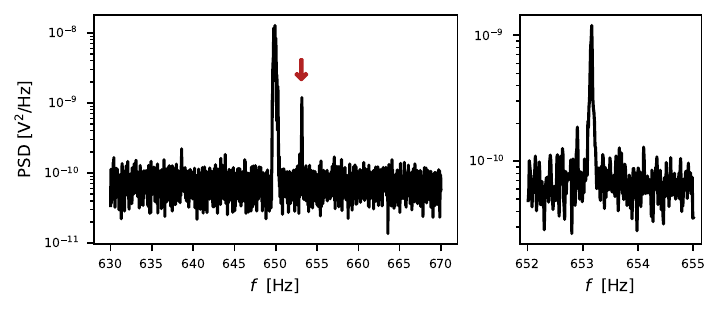}
\caption{Example of a thermal peak in the PSD of the SQUID readout signal. The thermal peak at $\approx653$~Hz is marked by a red arrow in the left panel and shown in greater detail in the right panel. The larger peak at 650~Hz is an electrical interference peak. The analysis of this particular trace yields an rms cantilever displacement of $0.4$~nm and $T_\textrm{cantilever}\approx290$~mK. }\label{fig_supp:example_thermal_motion}
\end{figure}

The analysis of the remaining 922 time traces focused on the frequency range between 652~Hz and 655~Hz, approximately centered around the cantilever resonant frequency $\omega_c\approx 653$ Hz.
In this small bandwidth, the PSD traces typically display a Lorentzian peak centered at $\omega_c$, which can be attributed to the thermal excitation of the cantilever oscillations.
In Fig.~\ref{fig_supp:example_thermal_motion} we show one illustrative example of such a thermal peak in the raw data.

The average energy stored in the oscillator during the measurement is proportional to the area of this thermal peak in the PSD.
To assign a value to the area on the basis of the noisy data, it is important to determine the noise background.
This was done separately for each trace, as follows.
First, we binned the data in six intervals of width $0.5$~Hz, such that the peak was contained in the third bin, while the remaining five bins only contained the noise.
We took as the noise power the average integral of these five bins, and subtracted it from the integral of the PSD in the bin containing the peak.
The result was first converted to an rms displacement of the oscillator, $\langle x^2 \rangle$, using the calibration procedure outlined in App.~\ref{app:calibration}, and then to the temperature of the oscillator via the equipartition theorem: $k_B T_\textrm{osc}=k\langle x^2\rangle$.

\end{appendices}

\bibliography{bibliography}

\end{document}